\documentclass{aastex61}
\usepackage{graphicx}
\usepackage{longtable}
\usepackage{multirow}
\usepackage{amsmath}
\usepackage{url}
\usepackage{amsmath,bm}
\usepackage{lipsum}
\usepackage{rotating}

\date{\today}

\begin{document}

\title{Hubble parameter and baryon acoustic oscillation measurement constraints on the Hubble constant, the deviation from the spatially flat $\Lambda$CDM model, the Deceleration-Acceleration transition redshift, and spatial curvature}

\author{Hai Yu}
\altaffiliation{yuhai@smail.nju.edu.cn}
\affiliation{School of Astronomy and Space Science, Nanjing University, Nanjing 210093, China}
\affiliation{Department of Physics, Kansas State University, 116 Cardwell Hall, Manhattan, KS 66506, USA}

\author{Bharat Ratra}
\altaffiliation{ratra@phys.ksu.edu}
\affiliation{Department of Physics, Kansas State University, 116 Cardwell Hall, Manhattan, KS 66506, USA}

\author{Fa-Yin Wang}
\altaffiliation{fayinwang@nju.edu.cn}
\affiliation{School of Astronomy and Space Science, Nanjing University, Nanjing 210093, China}
\affiliation{Key Laboratory of Modern Astronomy and Astrophysics (Nanjing University), Ministry of Education, Nanjing 210093, China}

\begin{abstract}
We compile a complete collection of reliable Hubble
parameter $H(z)$ data to redshift $z \leq 2.36$ and use them with the
Gaussian Process method to determine continuous $H(z)$ functions
for various data subsets. From these continuous $H(z)$'s, summarizing
across the data subsets considered, we find $H_0\sim 67 \pm 4\,\rm km/s/Mpc$,
more consistent with the recent lower values determined using a variety of
techniques. In most data subsets, we see a cosmological deceleration-acceleration
transition at 2$\sigma$ significance, with the data subsets transition
redshifts varying over $0.33<z_{\rm da}<1.0$ at 1$\sigma$ significance. We
find that the flat-$\Lambda$CDM model is consistent with the $H(z)$ data to
a $z$ of 1.5 to 2.0, depending on data subset considered, with 2$\sigma$
deviations from flat-$\Lambda$CDM above this redshift range. Using the
continuous $H(z)$ with baryon acoustic oscillation distance-redshift
observations, we constrain the current spatial curvature density parameter to
be $\Omega_{K0}=-0.03\pm0.21$, consistent with a flat universe, but the
large error bar does not rule out small values of spatial curvature that
are now under debate.
\end{abstract}


\section{Introduction}\label{sec:intro}

In the standard spatially flat $\Lambda$CDM cosmological model
\citep{Peebles1984}, the cosmological constant $\Lambda$ dominates the current
cosmological energy budget and powers the currently accelerating cosmological
expansion. Cold dark matter (CDM) and baryonic matter are the second and third
biggest contributors to the energy budget now, followed by small contributions
from neutrinos and photons. Earlier on, non-relativistic (cold dark and
baryonic) matter dominated and drove the decelerating cosmological expansion.
For reviews of this scenario, see \citet{RatraVogeley2008, Martin2012, HutererShafer2018}. Many different observations are largely consistent with the standard
picture, including cosmic microwave background (CMB) anisotropies
\citep{PlanckCollaboration2016}, baryon acoustic oscillation (BAO) peak
position data \citep{Alametal2017}, Hubble parameter measurements
\citep{Farooqetal2017}, and Type Ia supernovae observations
\citep{Bertouleetal2014}, but there is still room for some dark energy dynamics
or spatial curvature, among other possibilities.

In this context, the Hubble parameter data, the cosmological expansion rate
as a function of redshift $z$, $H(z)$, is particularly interesting. Current
$H(z)$ data covers a large redshift range, $0.07\leq z \leq 2.36$ \citep{Simonetal2005,Sternetal2010, Morescoetal2012,Blakeetal2012, Font-Riberaetal2014,Delubacetal2015,Moresco2015,Morescoetal2016, Alametal2017}, larger than that covered
by the Type Ia supernovae. Not only can the $H(z)$ data be used to constrain
the usual cosmological parameters, such as the non-relativistic matter and dark
energy density parameters \citep{SamushiaRatra2006},\footnote{Early developments
include \citet{Samushiaetal2007, ChenRatra2011b, Farooqetal2013a} while more
recent work may be traced back through \citet{Tripathietal2017, Lonappanetal2017, Rezaeietal2017, Maganaetal2017,AnagnostopoulosBasilakos2017, Martins2017}.}
they can also be used to trace the cosmological deceleration-acceleration
transition \citep{FarooqRatra2013,Jesusetal2017}, and be used to measure the Hubble constant
$H_0$ \citep{Bustietal2014, Chenetal2017, WangMeng2017}. And in conjunction
with distance-redshift data, $H(z)$ measurements can be used to constrain
spatial curvature \citep{Clarksonetal2007, Clarksonetal2008}.

In this paper, we build on and extend these results. We first gather together a
complete collection of currently available $H(z)$ data, which does not include
older results from earlier analyses of data subsets nor estimates
that are no longer believed to be reliable. Some of these measurements are
correlated, so we account for these correlations when using the Gaussian
Process (GP) method to determine a continuous $H(z)$ function that best
approximates the discrete $H(z)$ measurements we have collected. As far as
we are aware, such correlations have been ignored in all previous GP method
$H(z)$ determinations. (The technique we have developed here for accounting
for such correlations will be useful for future analyses.) From the GP method-determined $H(z)$ we measure $H_0$ and find it is more consistent with other lower $H_0$ estimates \citep{ChenRatra2011a}, but the high local measurement
\citep{Riessetal2016} lies within the 2$\sigma$ range of our estimates here.

For the first time, we use the evolution of $\Omega_{m0}$ with redshift
(derived from continuous $H(z)$ functions determined by applying the GP 
method to the observed $H(z)$ data) to test the spatially flat
$\Lambda$CDM model, finding agreement up to a $z$ of 1.5 or 2, depending on
which combination of discrete $H(z)$ data we use to determine the GP method
continuous $H(z)$. Above a redshift of 1.5 or 2, there are 2$\sigma$ indications
of deviation from flat-$\Lambda$CDM, but it has only recently become possible
to measure $H(z)$'s at or above $z=2$ so it is perhaps best to wait for better
and more higher-$z$ data before forming a strong opinion about these
deviations.

We also use these $H(z)$'s to study the cosmological deceleration-acceleration
transition and measure the redshift of this transition, $z_{\rm da}$. Our results
here qualitatively agree with those found earlier, with the GP method $H(z)$
constraints on $z_{\rm da}$ being significantly weaker than those determined by
using cosmological model templates \citep{Farooqetal2017}.

When the open inflation \citep{Gott1982, RatraPeebles1994, RatraPeebles1995}
and closed inflation \citep{Hawking1984, Ratra1985, Ratra2017} model
energy density inhomogeneity power spectra are used to analyze the Planck
2015 CMB anisotropy data \citep{PlanckCollaboration2016}, they favor a
closed universe with current spatial curvature density parameter magnitude of a percent
or two \citep{Oobaetal2017a, Oobaetal2017b, Oobaetal2017c, ParkRatra2018}. It is important to measure spatial
curvature in as many other ways as possible so as to definitely establish
whether the universe is spatially flat or spatially curved. The GP method-determined continuous $H(z)$ can be used, in conjunction with BAO measured
distance as function of redshift, to constrain the current value of the
spatial curvature energy density parameter $\Omega_{K0}$ \citep{Clarksonetal2007, Clarksonetal2008}. Here, we generalize the method proposed by
\citet{YuWang2016} for
measuring $\Omega_{K0}$, by now also accounting --- for the first time --- for
the correlations between some of the BAO distance measurements. (The technique
we have developed here for accounting for such correlations will be useful for
future analyses.) We find that
the data are consistent with a flat universe, with $\Omega_{K0}$ error bars
of about 0.2, much too large to test the findings of \citet{Oobaetal2017a, Oobaetal2017b, Oobaetal2017c}, and \citet{ParkRatra2018}.

Our paper is organized as follows. In the next section, we introduce our $H(z)$
data compilation and discuss how we organize these data into 12 different
samples. In Sec.\ \ref{sec:method}, we use the GP method to compute continuous
$H(z)$ functions that best represent our 12 samples of discrete $H(z)$ data.
In this section, we also measure $H_0$ from each of these samples. In Sec.\
\ref{sec:dOm}, we use the continuous $H(z)$ functions to test whether the
current $H(z)$ data are consistent with the flat-$\Lambda$CDM model or not, and
in Sec.\ \ref{sec:constrain1}, we measure the cosmological
deceleration-acceleration transition redshift $z_{\rm da}$ for each sample.
In Sec.\ \ref{sec:constrain2}, we gather the best available BAO measurements
and use them in a joint analysis with one of the continuous $H(z)$'s to
constrain $\Omega_{K0}$. We conclude in Sec.\ \ref{sec:summary}.

\section{Hubble parameter data}\label{sec:Hzdata}

The Hubble parameter measurements we use are taken from Table 1 of
\citet{Farooqetal2017} with the following alterations. We now also include
the recent redshift $z=0.4$ cosmic chronometric measurement \citep{Ratsimbazafyetal2017}. We drop the three \citet{Blakeetal2012} WiggleZ radial BAO points ---
because of the partial overlap of the WiggleZ and BOSS spatial regions
\citep{Beutleretal2016} that is difficult to account for in our analyses ---
choosing to instead retain the more precise BOSS radial BAO measurements
\citep{Alametal2017}. We rescale the five BAO $H(z)$ measurements in our
compilation to a fiducial sound horizon length
$r_{d,{\rm fid}} = 147.60 \pm 0.43$ Mpc determined from
the TT + lowP + lensing Planck 2015 analysis \citep{PlanckCollaboration2016}.
This results in 36 $H(z)$
measurements that are tabulated in Table \ref{tab:H_z}. This is a complete
collection of currently available, reliable $H(z)$ data.

\begin{table}
\begin{center}
\caption{Hubble parameter data.}\label{tab:H_z}
\begin{tabular}{l|c|c|c}
  \hline
  \hline
  % after \\: \hline or \cline{col1-col2} \cline{col3-col4} ... \cline{col4-col5} ...
$z$ &   $H(z) [\rm\,km/s/Mpc]$    &  Method & Reference \\ \hline
0.07	&	69	$\pm$	19.6	& a & 	(1) \\
0.09	&	69	$\pm$	12	& a &	(2) \\
0.12	&	68.6	$\pm$	26.2	& a &	(1) \\
0.17	&	83	$\pm$	8	& a &	(2) \\
0.179	&	75	$\pm$	4	& a &	(3) \\
0.199	&	75	$\pm$	5	& a &	(3) \\
0.2	        &	72.9	$\pm$	29.6	& a &   (1) \\
0.27	&	77	$\pm$	14	& a &   (2) \\
0.28	&	88.8	$\pm$	36.6	& a &   (1) \\
0.352	&	83	$\pm$	14	& a &   (3) \\
0.38	&	81.9	$\pm$	1.9	& b &   (4) \\
0.3802	&	83	$\pm$	13.5	& a &   (5) \\
0.4	        &	95	$\pm$	17	& a &   (2) \\
0.4004	&	77	$\pm$	10.2	& a &   (5) \\
0.4247	&	87.1	$\pm$	11.2	& a &   (5) \\
0.4497	&	92.8	$\pm$	12.9	& a &   (5) \\
0.47        &       89      $\pm$   50      & a &   (6) \\
0.4783	&	80.9	$\pm$	9	& a &   (5) \\
0.48	&	97	$\pm$	62	& a &   (6) \\
0.51	&	90.8	$\pm$	1.9	& b &   (4) \\
0.593	&	104	$\pm$	13	& a &   (3) \\
0.61	&	97.8	$\pm$	2.1	& b &   (4) \\
0.68	&	92	$\pm$	8	& a &   (3) \\
0.781	&	105	$\pm$	12	& a &   (3) \\
0.875	&	125	$\pm$	17	& a &   (3) \\
0.88	&	90	$\pm$	40	& a &   (6) \\
0.9	        &	117	$\pm$	23	& a &   (2) \\
1.037	&	154	$\pm$	20	& a &   (3) \\
1.3	        &	168	$\pm$	17	& a &   (2) \\
1.363	&	160	$\pm$	33.6	& a &   (8) \\
1.43	&	177	$\pm$	18	& a &   (2) \\
1.53	&	140	$\pm$	14	& a &   (2) \\
1.75	&	202	$\pm$	40	& a &   (2) \\
1.965	&	186.5	$\pm$	50.4	& a &   (8) \\
2.34	&	223	$\pm$	7	& c &   (9) \\
2.36	&	227	$\pm$	8	& c &   (10) \\ \hline
\end{tabular}
\begin{flushleft}
{Notes:
a. Cosmic chronometric method.
b. BAO signal in galaxy distribution.
c. BAO signal in Ly$\alpha$ forest distribution alone, or cross-correlated
with QSOs.
Reference:
(1).\ \cite{Zhangetal2014}, (2).\ \cite{Simonetal2005},
(3).\ \cite{Morescoetal2012}, (4).\ \cite{Alametal2017},
(5).\ \cite{Morescoetal2016}, (6).\ \cite{Ratsimbazafyetal2017},
(7).\ \cite{Sternetal2010}, (8).\ \cite{Moresco2015},
(9).\ \cite{Delubacetal2015}, (10).\ \cite{Font-Riberaetal2014}.
}
\end{flushleft}
\end{center}
\end{table}

Of these 36 $H(z)$ measurements, 31 are determined using the cosmic chronometric
technique\footnote{As discussed in \citet{Morescoetal2016}, see their Table 3 for example, cosmic chronometric $H(z)$ error bars are dominated by systematic uncertainty.}, three correlated measurements are from the radial BAO signal in the
galaxy distribution, and the last two at $z=2.34$ and $2.36$ are measured from
the BAO signal in the Ly$\alpha$ forest distribution alone or cross-correlated with QSOs. The covariance matrix of the three galaxy distribution
radial BAO $H(z)$ measurements is \citep{Alametal2017}
\begin{equation}\label{eq:matrix1}
 \left(
   \begin{array}{ccc}
3.65	&	1.78	&	0.93	\\
1.78	&	3.65	&	2.20	\\
0.93	&	2.20	&	4.45	\\
   \end{array}
 \right)
\end{equation}
and is accounted for in our computations here.

In addition, we also consider two different values for $H_0$,
$68.0\pm2.8\rm\,km/s/Mpc$ \citep{ChenRatra2011a} and
$73.24\pm1.74\rm\,km/s/Mpc$ \citep{Riessetal2016}, to study the effect of
the assumed $H_0$ value on our results.

Given that BAO results depend on the assumed cosmological model used to analyze
the BAO data (this is likely to be a small effect since there is fairly strong
evidence that the true cosmological model cannot be very different from the
models assumed for these analyses), it is useful to examine a variety of
different combinations of the $H(z)$ data.

Our Sample 1{\_}0 comprises the 31 cosmic chronometric measurements and the two highest redshift Ly$\alpha$ measurements, with labels a and c in column 3 of Table \ref{tab:H_z}, for a total of 33 points. Sample 2{\_}0 is the full collection of 36 measurements in Table \ref{tab:H_z}. Sample 3{\_}0 consists of only the 31 cosmic chronometric measurements which are labeled a in column 3 of Table \ref{tab:H_z}. Sample 4{\_}0  adds to these the three \citet{Alametal2017} BAO measurements, labeled b in column 3 of Table \ref{tab:H_z}, for a total of 34 points.

We add $H_0 = 68.0\pm2.8\rm\,km/s/Mpc$ as the prior value of the Hubble constant to the samples above and denote this new set as Samples 1{\_}1, 2{\_}1, 3{\_}1, and 4{\_}1, with 34, 37, 32, and 35 data points respectively. In order to study the effect of the choice of the prior $H_0$ value, we also consider  Samples 1{\_}2, 2{\_}2, 3{\_}2, and 4{\_}2 which instead use $H_0 = 73.24\pm1.74\rm\,km/s/Mpc$ as the prior value. All in all, we consider 12 different $H(z)$ samples.

\section{Smoothed $H(z)$ function from the Gaussian Process method}\label{sec:method}

To leverage the $H(z)$ data, it is necessary to assume a cosmological model
that is characterized in terms of a small number of free parameters and to
use the $H(z)$ data to constrain these free parameters. $\Lambda$CDM \citep{Peebles1984} and $\phi$CDM \citep{PeeblesRatra1988, RatraPeebles1988} are two
physically consistent dark energy (either $\Lambda$ or
a dynamical scalar field $\phi$) CDM models that have been
used for this purpose. Often, a parameterization, XCDM, in which dynamical dark
energy is modeled as an ideal fluid,
has also been used, but this parameterization is physically inconsistent
and does not adequately approximate the dark energy evolution of $\phi$CDM
\citep{PodariuRatra2001}.

Here, we use the GP method to determine a continuous
function $H(z)$ that best represents the discrete Hubble parameter data we
have compiled in Table
\ref{tab:H_z}.\footnote{Other methods have also been used to determine
continuous functions that best represent discrete cosmological data \citep{MignoneBartelmann2008, MaturiMignone2009, Benitez-Herreraetal2012, Montieletal2014, VitentiPenna-Lima2015,SemizCamlibel2015}.} The GP method was first used cosmologically by \citet{Holsclawetal2010a,Holsclawetal2010b, Holsclawetal2011, Shafielooetal2012, Seikeletal2012a, Seikeletal2012b}. From a continuous $H(z)$, we are able to extract interesting cosmological information,
including the values of $H_0$ and $z_{\rm da}$.

\subsection{Gaussian Process method}\label{ssec:GP}

The GP method is used to obtain a continuous function $f(x)$ that is best
representative of a discrete set of measurements $f(x_i)\pm\sigma_i$ at $x_i$,
where $i = 1,2,...,N$ and $\sigma_i$ are the 1$\sigma$ error bars. The GP
method assumes that the value of the function at any position $x$ is a random variable
that follows a gaussian distribution. And the expectation and standard deviation of
this gaussian distribution, $\mu(x)$ and $\sigma(x)$, are determined from the
discrete data through a defined covariance function (or kernel function)
$k(x,x_i)$ and are given by
\begin{equation}\label{eq:mu}
    \mu(x)=\sum_{i,j=1}^Nk(x,x_i)(M^{-1})_{ij}f(x_j),
\end{equation}
and
\begin{equation}\label{eq:sigma}
    \sigma(x)=k(x,x)-\sum_{i,j=1}^Nk(x,x_i)(M^{-1})_{ij}k(x_j,x),
\end{equation}
where the matrix $M_{ij}=k(x_i,x_j)+c_{ij}$ and $c_{ij}$ is the covariance
matrix of the observed data, given by eqn.\ (\ref{eq:matrix1}) for the
correlated measurements and otherwise diagonal with elements $\sigma_i^2$.
Equations (\ref{eq:mu}) and (\ref{eq:sigma}) specify the posterior distribution
of the extrapolated points.

Given eqns.\ (\ref{eq:mu}) and (\ref{eq:sigma}), the continuous function
$f(x)$ can be determined once we have a suitable covariance function
$k(x,x^\prime)$. In practice, there are many possible covariance functions.
Here, we consider three covariance functions to illustrate the ``model
dependence'' of our results. The usual covariance function, and that used
in most of our analyses here, is the gaussian
\begin{equation}\label{eq:sqex}
    k(x,x^\prime)=\sigma_f^2\exp{\left[-\frac{(x-x^\prime)^2}{2l^2}\right]}.
\end{equation}
Here, $\sigma_f$ and $l$ are parameters that control the strength of
the correlation of the function value and the coherence length of the
correlation in $x$, respectively.
The other two covariance functions we use to examine the ``model dependence''
of our results are the Mat$\rm\acute{e}$rn and Cauchy ones. The forms of
these are
\begin{equation}\label{eq:Matern}
    k(x,x^\prime)=\sigma_f^2 \left[1+\frac{\sqrt{3}|x-x^\prime|}{l}\right]\exp{\left[-\frac{\sqrt{3}|x-x^\prime|}{l}\right]},
\end{equation}
and
\begin{equation}\label{eq:Cauchy}
    k(x,x^\prime)=\sigma_f^2\frac{l}{(x-x^\prime)^2+l^2}.
\end{equation}
(We find good consistency between the continuous functions we derive using the three different covariance functions.)
The parameters $\sigma_f$ and $l$ are optimized for the observed data,
$f(x_i)\pm\sigma_i$, by minimizing the log marginal likelihood function \citep{Seikeletal2012a}
\begin{equation}\label{likelihood}
  \ln\mathcal{L} = -\frac{1}{2}\sum_{i,j=1}^N[f(x_i)-\mu(x_i)](M^{-1})_{ij}[f(x_j)-\mu(x_j)]-\frac{1}{2}
   \ln|M|-\frac{1}{2}N\ln{2\pi},
\end{equation}
where $|M|$ is the determinant of $M_{ij}$. 

In this work, we use the open-source Python package GaPP
\citep{Seikeletal2012a}, which is widely used in cosmological studies
\citep{Seikeletal2012b, BilickiSeikel2012, Caietal2016, Wangetal2017a, YuWang2017}. It can output the continuous function $f(x)$ as well as its first derivative $f'(x)$ once certain data and parameters are input. The derivative is computed from the smooth reconstructed function and its uncertainty is estimated from the covariance function. For detailed information about the Gaussian Process method and the package GaPP, see \cite{Seikeletal2012a} and www.acgc.uct.ac.za/\~{}seikel/GAPP/Documentation/Documentation.html.

\subsection{Smoothed $H(z)$ function and $H_0$}\label{ssec:H(z)}

%\begin{sidewaysfigure}
\begin{figure}
  \includegraphics[width=1.3\textwidth,angle=90]{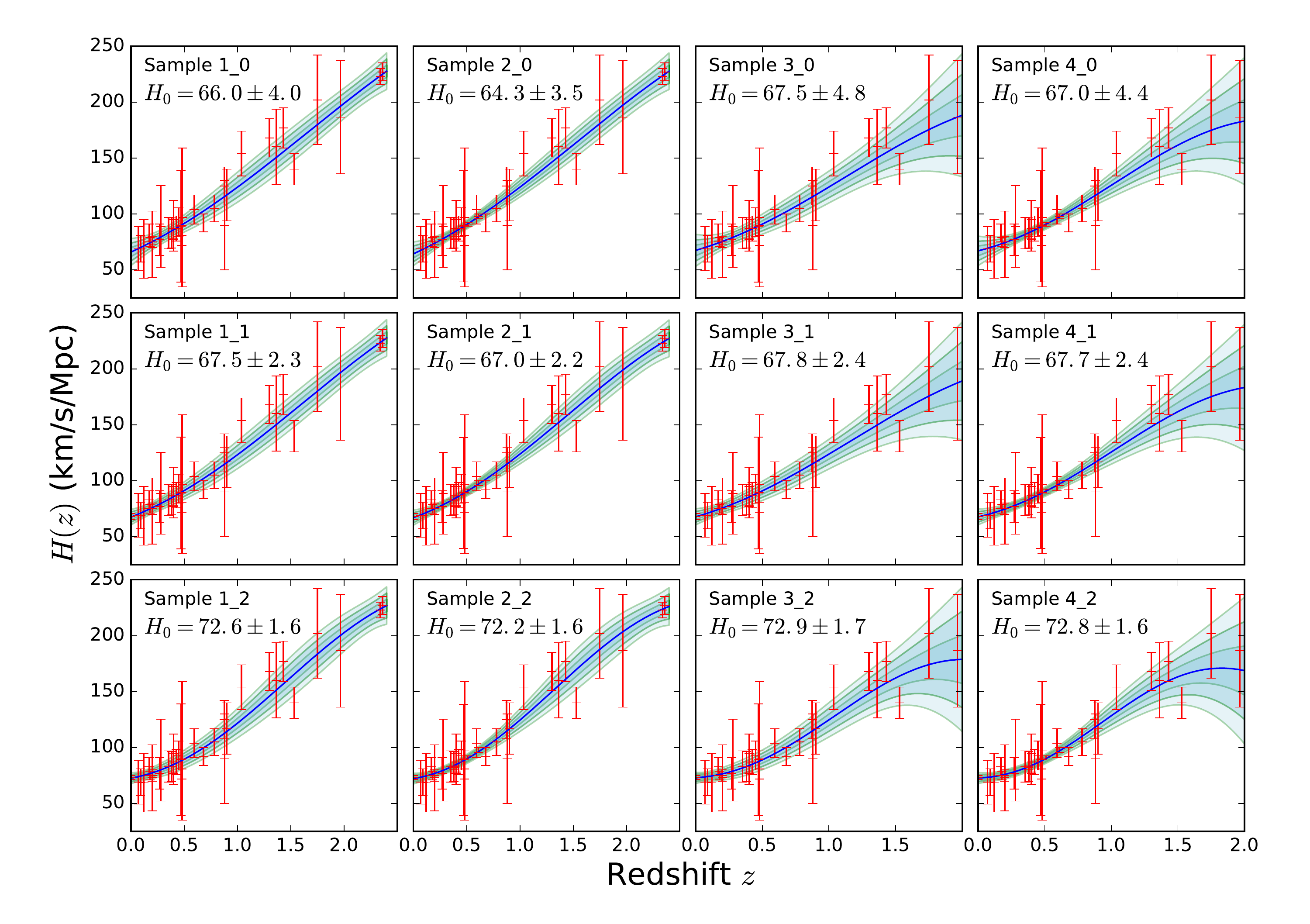}\\
  \caption{Smoothed $H(z)$ functions for all 12 samples. The blue lines are
the mean curves and the shadow areas are $1\sigma$, $2\sigma$, and $3\sigma$
confidence regions.}\label{fig:H(z)}
\end{figure}
%\end{sidewaysfigure}

In this section, we use the 12 samples of Sec.\ \ref{sec:Hzdata} and the GP
method\footnote{We note that lower $z$ parts of earlier compilations of
$H(z)$ data have been
shown to not be inconsistent with Gaussianity and so probably consistent
with the GP method requirement of Gaussianity \citep{Farooqetal2013b, Farooqetal2017}.}
summarized in the previous sub-section to compute continuous Hubble
parameter functions $H(z)$ and use them to constrain the value of the
Hubble constant $H_0$.\footnote{For earlier studies based on smaller
$H(z)$ compilations, see \citet{Bustietal2014, WangMeng2017}.}
The smoothed $H(z)$ functions from all 12 samples
are plotted in Figure\ \ref{fig:H(z)}. The three panels of the first column
are the results from Samples 1\_0, 1\_1, and 1\_2, respectively and the other
three columns are similar but for Sample 2, 3, and 4, respectively. From
these $H(z)$, we can measure $H_0$; these are listed in Table \ref{tab:H0}.

Table \ref{tab:H0} also lists $H_0$ values derived under the
Mat$\rm\acute{e}$rn and Cauchy  covariance function assumptions, in addition to
those derived using the gaussian covariance function (which is used to
compute the $H(z)$ functions shown in Figure\ \ref{fig:H(z)}). We also plot all of the $H_0$ values together (see the left panel of Figure\ \ref{fig:Hz-z_da}). It is reassuring
that changing the covariance function used does not significantly alter the
estimated $H_0$, even more so when we use an $H_0$ measurement in the
determination of $H(z)$.

Comparing the three panels in each column in Figure\ \ref{fig:H(z)}, or the
equivalent determined $H_0$ values listed in Table \ref{tab:H0}, we see that
if one of the $H_0$ measurements is used in the determination of the continuous
$H(z)$, then the resulting errors on the determined $H_0$ are smaller. This is
because the $H_0$ measurements have much tighter error bars than the rest of
the $H(z)$ data we use.

The first row of Figure\ \ref{fig:H(z)}, or the corresponding $H_0$ values
listed in Table \ref{tab:H0}, show that all four of the non-$H_0$ samples
we examine (Samples 1{\_}0, 2{\_}0, 3{\_}0, and 4{\_}0) favor a lower value
of $H_0$, but with fairly large error bars (about $\pm 4\,\rm km/s/Mpc$).
These are consistent with the most recent median statistics estimate
$H_0=68\pm2.8\,\rm km/s/Mpc$ \citep{ChenRatra2011a}, which is consistent
with earlier median statistics estimates \citep{Gottetal2001, Chenetal2003}.
These values are also quite consistent with cosmological-model-based
determinations of $H_0$ from $H(z)$ data \citep{Chenetal2017}. Many recent
estimates of $H_0$ are also quite consistent with the median statistics
measurement \citep{Calabreseetal2012, Sieversetal2013, Aubourgetal2015, SemizCamlibel2015, PlanckCollaboration2016, LHuillierShafieloo2017, Lukovicetal2016, Wangetal2017b, LinIshak2017, Abbottetal2017}. Of course, the error
bars on $H_0$ estimated here and in \citet{Chenetal2017} are large and the
high local measurement of $H_0 = 73.24\pm1.74\rm\,km/s/Mpc$
\citep{Riessetal2016} lies within the 2$\sigma$ confidence limits of our
measurements here. In addition, we note that some other local expansion rate
measurements find a slightly lower $H_0$ with larger error bars \citep{Rigaultetal2015, Zhangetal2017, Dhawanetal2018, FernandezArenasetal2017}.

\begin{table}
\begin{center}
\caption{Constraints on $H_0$ and $z_{\rm da}$ from 12 samples and 3 different covariance functions.}\label{tab:H0}
\begin{tabular}{c|c|c|c}
\hline \hline
Sample & Covariance & $H_0$ & $z_{\rm da}$\\
ID &  Function & $(\rm km/s/Mpc)$ & \\ \hline

	&	G	&	66.0	$\pm$	4.0	& $	0.59	^{+	0.19	}_{-	0.32	 }$ \\
Sample 1\_0	&	M	&	67.8	$\pm$	5.3	& $	0.57	^{+	0.27	 }_{-	 0.23	}$ \\
	&	C	&	66.9	$\pm$	4.2	& $	0.59	^{+	0.18	}_{-	0.24	 }$ \\ \hline
	&	G	&	67.5	$\pm$	2.3	& $	0.63	^{+	0.14	}_{-	0.14	 }$ \\
Sample 1\_1	&	M	&	67.9	$\pm$	2.5	& $	0.57	^{+	0.26	 }_{-	 0.19	}$ \\
	&	C	&	67.7	$\pm$	2.3	& $	0.61	^{+	0.16	}_{-	0.13	 }$ \\ \hline
	&	G	&	72.6	$\pm$	1.6	& $	0.65	^{+	0.11	}_{-	0.08	 }$ \\
Sample 1\_2	&	M	&	72.8	$\pm$	1.7	& $	0.57	^{+	0.22	 }_{-	 0.15	}$ \\
	&	C	&	72.7	$\pm$	1.6	& $	0.63	^{+	0.12	}_{-	0.09	 }$ \\ \hline
	&	G	&	64.3	$\pm$	3.5	& $	0.50	^{+	0.17	}_{-	0.25	 }$ \\
Sample 2\_0	&	M	&	66.4	$\pm$	4.8	& $	0.46	^{+	0.32	 }_{-	 0.12	}$ \\
	&	C	&	65.0	$\pm$	3.7	& $	0.51	^{+	0.16	}_{-	0.21	 }$ \\ \hline
	&	G	&	67.0	$\pm$	2.2	& $	0.55	^{+	0.12	}_{-	0.10	 }$ \\
Sample 2\_1	&	M	&	67.6	$\pm$	2.4	& $	0.47	^{+	0.31	 }_{-	 0.10	}$ \\
	&	C	&	67.1	$\pm$	2.3	& $	0.55	^{+	0.12	}_{-	0.10	 }$ \\ \hline
	&	G	&	72.2	$\pm$	1.6	& $	0.58	^{+	0.09	}_{-	0.07	 }$ \\
Sample 2\_2	&	M	&	72.7	$\pm$	1.7	& $	0.46	^{+	0.34	 }_{-	 0.08	}$ \\
	&	C	&	72.3	$\pm$	1.6	& $	0.56	^{+	0.09	}_{-	0.07	 }$ \\ \hline
	&	G	&	67.5	$\pm$	4.8	& $	0.56	^{+	0.22	}_{-	0.25	 }$ \\
Sample 3\_0	&	M	&	68.8	$\pm$	6.3	& $	0.55	^{+	0.23	 }_{-	 0.20	}$ \\
	&	C	&	69.6	$\pm$	5.2	& $	0.53	^{+	0.16	}_{-	0.14	 }$ \\ \hline
	&	G	&	67.8	$\pm$	2.4	& $	0.58	^{+	0.23	}_{-	0.13	 }$ \\
Sample 3\_1	&	M	&	68.1	$\pm$	2.5	& $	0.55	^{+	0.45	 }_{-	 0.19	}$ \\
	&	C	&	68.1	$\pm$	2.4	& $	0.55	^{+	0.20	}_{-	0.12	 }$ \\ \hline
	&	G	&	72.9	$\pm$	1.7	& $	0.55	^{+	0.12	}_{-	0.08	 }$ \\
Sample 3\_2	&	M	&	72.9	$\pm$	1.7	& $	0.54	^{+	0.21	 }_{-	 0.16	}$ \\
	&	C	&	73.0	$\pm$	1.7	& $	0.53	^{+	0.12	}_{-	0.08	 }$ \\ \hline
	&	G	&	67.0	$\pm$	4.4	& $	0.47	^{+	0.12	}_{-	0.13	 }$ \\
Sample 4\_0	&	M	&	67.7	$\pm$	5.5	& $	0.44	^{+	0.56	 }_{-	 0.11	}$ \\
	&	C	&	68.1	$\pm$	4.7	& $	0.47	^{+	0.11	}_{-	0.10	 }$ \\ \hline
	&	G	&	67.7	$\pm$	2.4	& $	0.48	^{+	0.11	}_{-	0.08	 }$ \\
Sample 4\_1	&	M	&	67.9	$\pm$	2.5	& $	0.45	^{+	0.55	 }_{-	 0.10	}$ \\
	&	C	&	67.9	$\pm$	2.4	& $	0.47	^{+	0.11	}_{-	0.08	 }$ \\ \hline
	&	G	&	72.8	$\pm$	1.6	& $	0.48	^{+	0.08	}_{-	0.05	 }$ \\
Sample 4\_2	&	M	&	72.8	$\pm$	1.7	& $	0.44	^{+	0.56	 }_{-	 0.08	}$ \\
	&	C	&	72.8	$\pm$	1.6	& $	0.48	^{+	0.08	}_{-	0.05	 }$ \\ \hline

\end{tabular}
\begin{flushleft}
{Notes:
G: Gaussian covariance function,
M: Mat$\rm\acute{e}$rn covariance function,
C: Cauchy covariance function.
}
\end{flushleft}
\end{center}
\end{table}

\section{Testing the flat-$\Lambda$CDM model}\label{sec:dOm}

\begin{figure}
  \includegraphics[width=1.3\textwidth,angle=90]{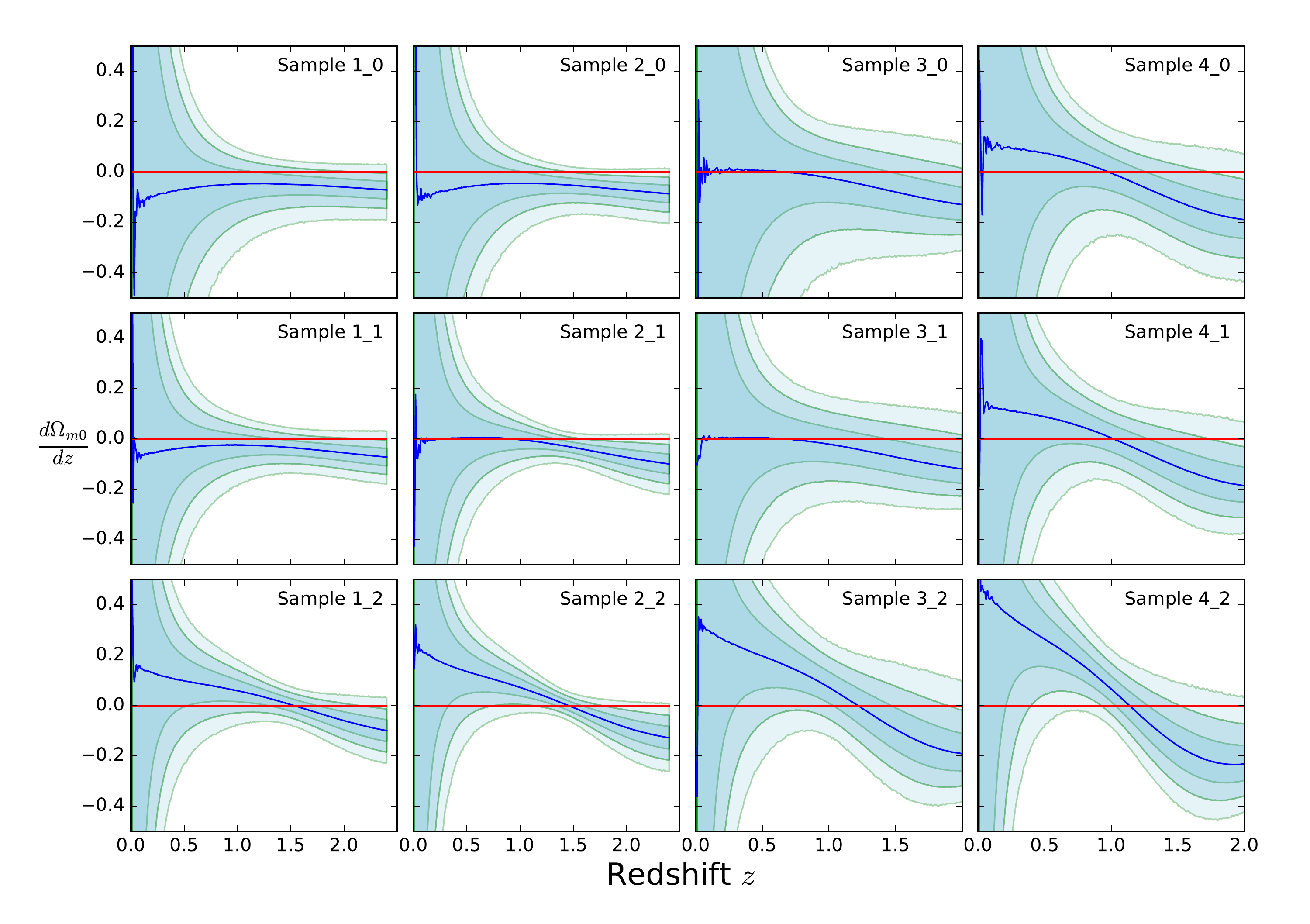}\\
  \caption{$\frac{d\Omega_{m0}}{dz}$ of all the 12 samples. The blue lines are the mean curves and the gray areas are $1\sigma$, $2\sigma$, and $3\sigma$ confidence regions respectively. The red lines indicate $\frac{d\Omega_{m0}}{dz}=0$.}\label{fig:dOm}
\end{figure}

In the spatially flat $\Lambda$CDM model, which provides a reasonable
description of the universe, the expansion history is
$H(z)=H_0\sqrt{\Omega_{m0}(1+z)^3+\Omega_{r0}(1+z)^4+1-\Omega_{m0}-\Omega_{r0}}$ where $\Omega_{m0}$ and $\Omega_{r0}$ are the current values of the non-relativistic and relativistic matter density parameters. Solving, we have
\begin{equation}
\frac{[H^2/H_0^2-1]}{\Omega_{m0}[(1+z)^3-1]}=1-\frac{\Omega_{r0}[(1+z)^4-1]}{\Omega_{m0}[(1+z)^3-1]} \approx 1,
\end{equation}
where $\frac{\Omega_{r0}[(1+z)^4-1]}{\Omega_{m0}[(1+z)^3-1]}\sim10^{-3}$ at $z = 2$ and can be omitted because it is much smaller than 1 in the redshift range we care about and the uncertainties of $H(z)$ data are relatively large. Therefore, we have
\begin{equation}\label{eq:Om}
    \Omega_{m0} = \left[\left(\frac{H(z)}{H_0}\right)^2-1\right]/[(1+z)^3-1],
\end{equation}
which is a redshift-independent constant if flat-$\Lambda$CDM is the
correct description of the universe and so can be used to test the model
\citep{Sahnietal2008}. To test whether this is constant, we compute the first
derivative of the right hand side of this equation, which we label
$\frac{d\Omega_{m0}}{dz}$. If this is significantly biased
away from 0, it implies that $\Omega_{m0}$ does evolve with redshift and that
the data don't favor the flat-$\Lambda$CDM model.\footnote{For an application
of this test based on supernovae data, see \citet{Yahyaetal2014}.}

The $\frac{d\Omega_{m0}}{dz}$ function can be derived from the smoothed $H(z)$
function and results for all 12 samples are shown in Figure\ \ref{fig:dOm}.
The $H(z)$ data are largely consistent with the flat-$\Lambda$CDM model, with
some deviations at higher $z$ for some of the samples. At 2$\sigma$, in column
3 of Figure\ \ref{fig:dOm}, for the 31 cosmic chronometric measurements alone or
in combination with an $H_0$ measurement, we see no deviation from
flat-$\Lambda$CDM except in the lowest panel above a $z$ of about 1.9.
Adding in the three \citet{Alametal2017} radial BAO measurements, column 4 of
Figure\ \ref{fig:dOm}, we see deviation from flat-$\Lambda$ above about a
redshift of 1.8 in the top and middle (with the added
$H_0=68\pm2.8\,\rm km/s/Mpc$ point) panels, and above about a $z$ of
1.6 in the lowest panel (with the added $H_0 = 73.24\pm1.74\rm\,km/s/Mpc$
point). Instead, when we consider the 31 cosmic chronometric measurements with
the two highest $z$ Ly$\alpha$ ones, column 1 of Figure\ \ref{fig:dOm}, the
trend is in the opposite direction, with the data being consistent with
flat-$\Lambda$CDM below a redshift of 1.9 or 2 for the top and middle panels,
while in the lowest panel flat-$\Lambda$CDM is adequate up to about a $z$
of 2.2. The full $H(z)$ data, in column 2, has a similar trend, with
flat-$\Lambda$CDM doing a reasonable job to a $z$ of about 1.6 in the top
two panels and to about 1.8 in the bottom panel.

So, depending on the sample considered, flat-$\Lambda$CDM provides an
adequate model to a redshift of 1.5 or 2, which is consistent with former a study \citep{Lonappanetal2017}. Higher $z$ $H(z)$ data is inconsistent with flat-$\Lambda$CDM at 2$\sigma$. However, given that it
has only recently become possible to make such high $z$ measurements, it is
probably best not to make too much of this disagreement. This is an
interesting test and we look forward to soon learning more about whether or not
flat-$\Lambda$CDM provides an adequate description of the $z\sim 2$ and higher
universe, when better quality and more $H(z)$ data becomes available.

\section{Constraining the cosmological deceleration-acceleration transition redshift}\label{sec:constrain1}

In the standard cosmological picture, dark energy dominates the current
cosmological energy budget and is responsible for the currently accelerating
cosmological expansion; at earlier times non-relativistic --- baryonic
and cold dark --- matter dominated the energy budget and powered the
decelerating expansion. \citet{FarooqRatra2013} used $H(z)$ data to study
this transition and measure the redshift of the transition, $z_{\rm da}$, in a
variety of cosmological models. For more recent similar analyses of
compilations of $H(z)$ measurements, see \citet{Capozzielloetal2014, Morescoetal2016, Farooqetal2017}. Here, we use the GP method $H(z)$ continuous functions
we have derived from the most up-to-date compilation of $H(z)$ data to measure
$z_{\rm da}$.

The Friedmann equation is
\begin{equation}\label{eq:ddota}
    \ddot{a}(z)=\frac{H^2}{1+z}-H\frac{dH}{dz},
\end{equation}
where $a$ is the scale factor of the universe. $z_{\rm da}$ is the solution of
$\ddot{a}(z_{\rm da})=0$. Using the continuous $H(z)$ functions obtained in
Sec.\ \ref{ssec:H(z)} and eqn.\ (\ref{eq:ddota}), we can derive $\ddot{a}(z)$
for each of the 12 samples and then solve for $z_{\rm da}$ from
$\ddot{a}(z_{\rm da})=0$.

\begin{figure}
  \includegraphics[width=1.3\textwidth,angle=90]{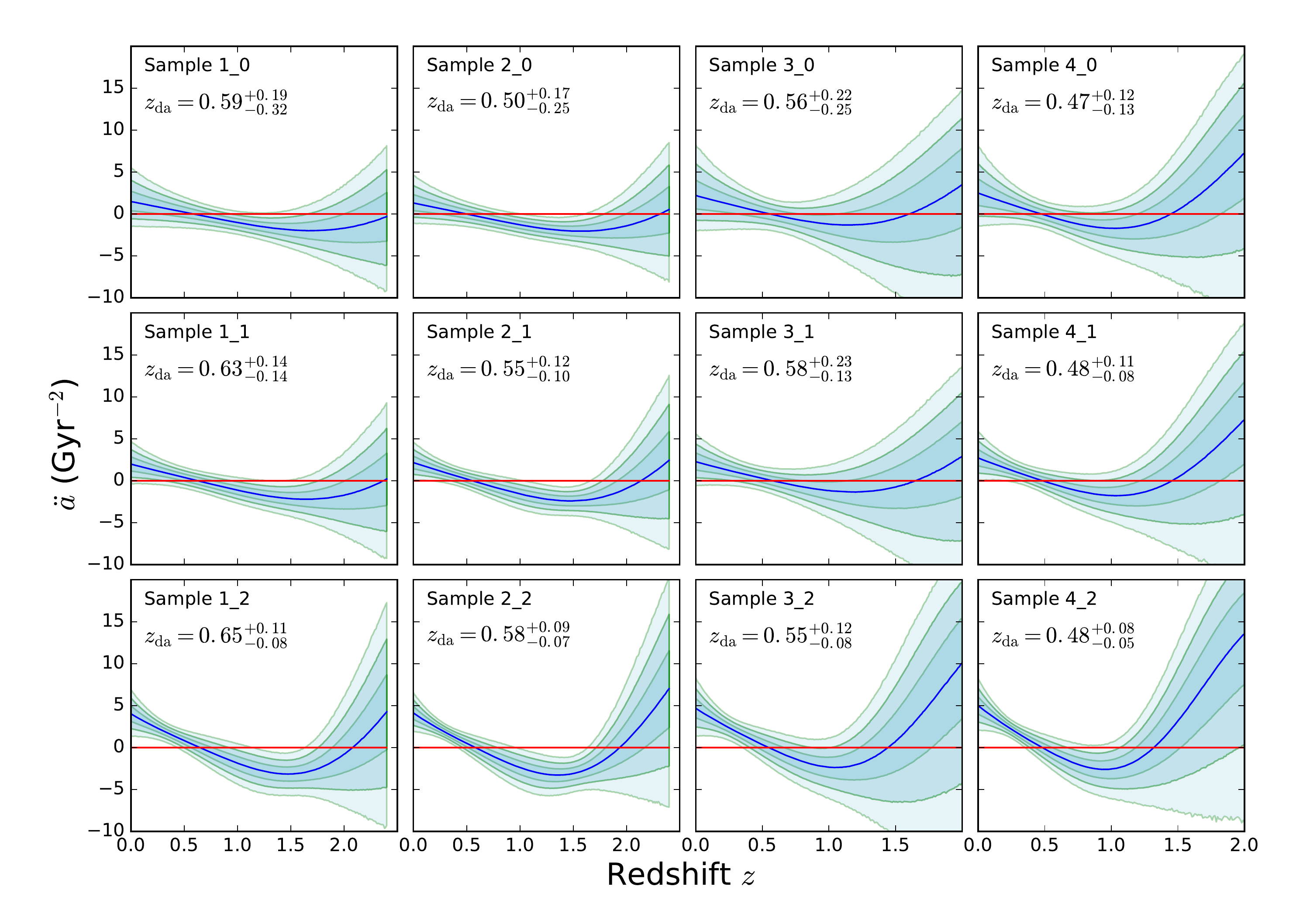}\\
  \caption{$\ddot{a}(z)$ for the 12 samples. The blue lines are the mean curves and the gray areas are $1\sigma$, $2\sigma$, and $3\sigma$ confidence regions, respectively. The red lines indicate $\ddot{a}=0$.}\label{fig:ddota}
\end{figure}

\begin{figure}
	\includegraphics[width=\textwidth]{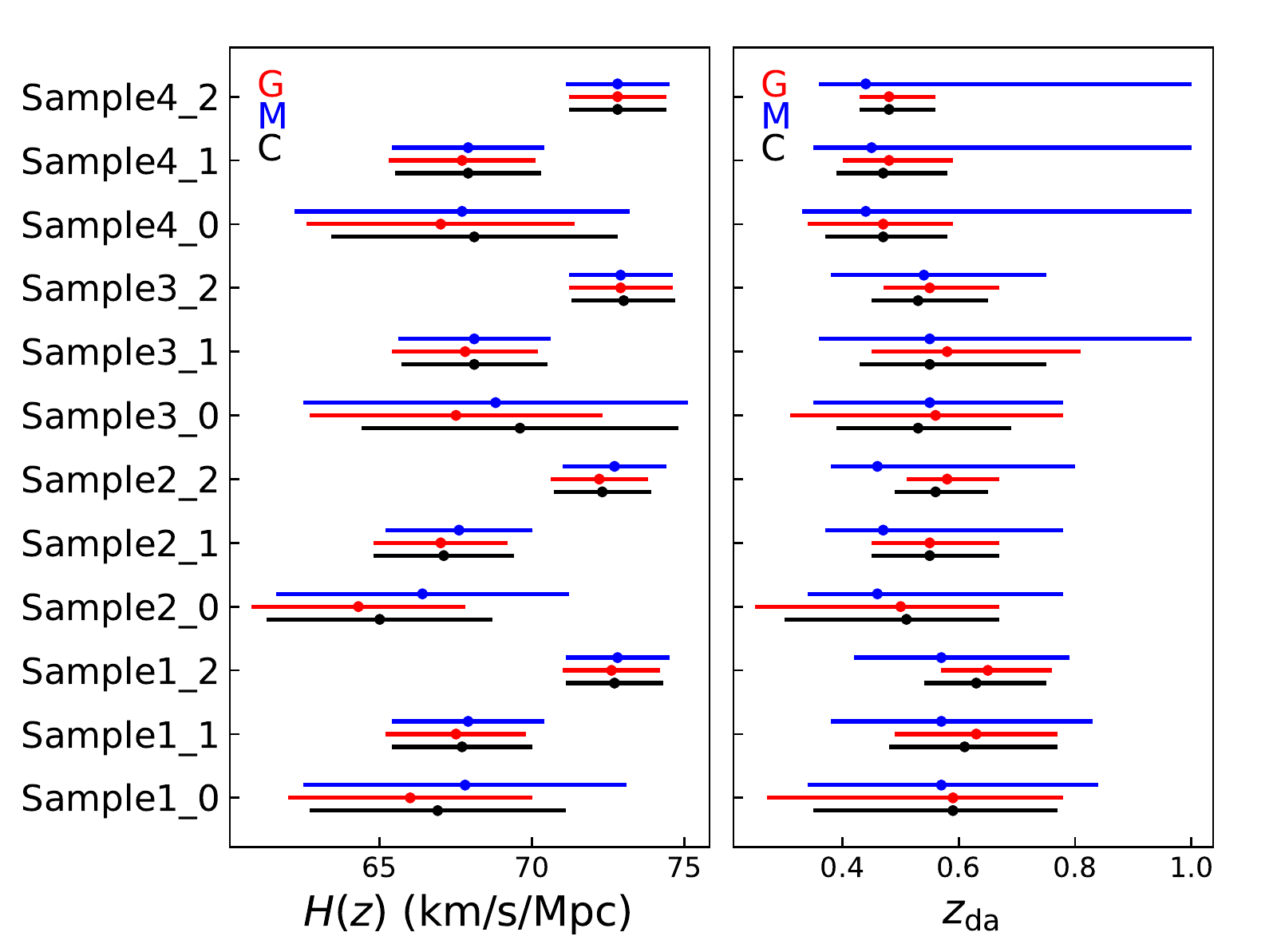}\\
	\caption{Determined $H_0$ and $z_{da}$ values for all 12 samples and three covariance functions. The errors are $1\,\sigma$ confidence level. The red, blue, and black colors represent the gaussian, Mat\'{e}rn, and Cauchy covariance function determinations.}\label{fig:Hz-z_da}
\end{figure}

Figure \ref{fig:ddota} shows the $\ddot{a}(z)$ functions for the 12 samples,
derived using the gaussian covariance function. From the functions plotted in the figure, we can read the values of $z_{\rm da}$, which occur where the (central) blue and red lines cross (at lower $z$). Similarly, the crossing points of the red line and the $1\,\sigma$ confidence region boundary determine the $1\,\sigma$ errors on $z_{\rm da}$. The last column of Table
\ref{tab:H0} lists the lowest $z_{\rm da}$ value that solves the corresponding
$\ddot{a}(z_{\rm da})=0$ for the 12 samples for each of the three covariance
functions we use. We also plot all of the $z_{\rm da}$ values on a same figure for easy comparison (see the right panel of Figure\ \ref{fig:Hz-z_da}). 
All of the determined values are largely consistent with each other no matter which sample or covariance function is used.
From the listed $z_{\rm da}$ and the figure, we see that, depending on sample,
there is a clear detection of a deceleration-acceleration transition in the
1$\sigma$ redshift range $0.33 < z_{\rm da} < 1.0$. 
This is quite a broad range, with most of the
larger $z_{\rm da}$ 1$\sigma$ upper values coming from the Mat$\rm\acute{e}$rn
covariance function case. This range is significantly larger than those
determined by using cosmological models \citep{FarooqRatra2013, Capozzielloetal2014,SemizCamlibel2015, Morescoetal2016, Farooqetal2017} or a two-part piece-wise linear fit to
the $H(z)$ data \citep{Morescoetal2016}. This is probably because the GP method
has more freedom than the cosmological model templates or the two-part
piece-wise linear template. While, for a given covariance function, the GP method
has only two parameters, $\sigma_f$ and $l$, there is the additional freedom
of being allowed to alter the covariance function assumed, thus
resulting in a larger range for $z_{\rm da}$. It is likely, given the present
quality of the $H(z)$ data, that the cosmological model templates result in a
more reliable determination of $z_{\rm da}$ than that obtained from the GP method
here.

While there is some evidence in Figure\ \ref{fig:ddota} to suggest an early
(additional) epoch of accelerated cosmological expansion, at 2$\sigma$
confidence, this is only in the lowest panel of column 4, for Sample 4{\_}2,
above a redshift of about 1.9, for the 31 cosmic chronometric and three
\citet{Alametal2017} BAO measurements and including
$H_0 = 73.24\pm1.74\rm\,km/s/Mpc$. More and better higher-$z$ $H(z)$ data will be needed to resolve this issue.

In the top row of Figure\ \ref{fig:ddota}, we see that at 2$\sigma$ only
Sample 4{\_}0, 31 cosmic chronometric and three \citet{Alametal2017} BAO
measurements, shows evidence for low-$z$ acceleration and intermediate-$z$
deceleration, while Samples 1{\_}0 and 2{\_}0 show evidence only for
intermediate-$z$ deceleration with Sample 3{\_}0 consistent with $\ddot{a}(z)=0$ at 2$\sigma$.

When an $H_0$ measurement is included with the other $H(z)$ data, in rows
2 and 3 of Figure\ \ref{fig:ddota}, we see 2$\sigma$ evidence for low-$z$
acceleration and intermediate-$z$ deceleration in all panels, except
in Sample 3{\_}1 for the 31 cosmic chronometric measurements where there
is no 2$\sigma$ evidence for intermediate-$z$ deceleration.

All in all, the results of the $\ddot{a}(z_{\rm da})=0$ analyses here are
qualitatively consistent with those found earlier \citep{FarooqRatra2013, Capozzielloetal2014, Morescoetal2016, Farooqetal2017}. More and better $H(z)$ data
will allow the GP method to provide $z_{\rm da}$ estimates that can also be
quantitatively compared to those derived from cosmological model templates.

\section{Constraining spatial curvature}\label{sec:constrain2}

The GP method-determined $H(z)$, the cosmological expansion rate as a function
of redshift, may be used in conjunction with BAO observations, which give a
distance as a function of redshift, to measure spatial curvature \citep{Clarksonetal2007, Clarksonetal2008, MortsellJonsson2011, Saponeetal2014, Lietal2014, TakadaDore2015, Caietal2016, YuWang2016, LHuillierShafieloo2017, Lietal2016, WeiWu2017, Ranaetal2017, Wangetal2017b}. Here, we extend the test proposed by \citet{YuWang2016} to constrain spatial curvature, to now allow for correlations
between the BAO measured distances. This method measures the spatial
curvature density parameter $\Omega_{K0}$ by
using the relation between the co-moving radial and angular diameter distances
\begin{equation}\label{eq:dpdm}
\frac{H_0D_M(z)}{c}\sqrt{-\Omega_{K0}} =
\sin\left(\frac{H_0D_C(z)}{c}\sqrt{-\Omega_{K0}}\right).
\end{equation}
Here, $D_C(z)=\int_0^z\frac{cdz}{H(z)}$ is the co-moving radial distance,
$D_M(z)=(1+z)D_A(z)$ is the co-moving angular diameter distance, $c$ is the
speed of light, $D_A(z)$ is the angular diameter distance, and for the open case with positive  $\Omega_{K0}$, the $\sin(x)$ is replaced with $\sin(ix)=i\sinh(x)$. We shall also
have need for the angle-averaged distance $D_V(z) = [c z D_M^2(z)/H(z)]^{1/3}$.

In Table \ref{tab:BAO} we compile BAO distance measurements we use. We
rescale these to the baryon drag epoch fiducial sound horizon length
$r_{d, {\rm fid}}=147.60\pm0.43$ Mpc from the Planck 2015 TT + lowP + lensing
analysis \citep{PlanckCollaboration2016}.

\begin{table}
\begin{center}
\caption{BAO distance measurements used in this work.}\label{tab:BAO}
\begin{tabular}{c|c|c|c|c}
\hline \hline
Variable & Redshift & Original Value & Rescaled Value & References \\
    &   &   (Mpc)   &   (Mpc)   &   \\ \hline
$D_V$	&	0.106	&	$456\pm27$	&	$456\pm27$	&	 \cite{Beutleretal2011}\\ \hline
$D_V$	&	0.15	&	$(664\pm25)r_d/r_{d, {\rm fid}}$	&	$659\pm25$	&	 \cite{Rossetal2015} \\ \hline
	&	0.38	&	$(1518\pm22)r_d/r_{d, {\rm fid}}$	&	$1516\pm22$	&	 \\
$D_M$	&	0.51	&	$(1977\pm27)r_d/r_{d, {\rm fid}}$	&	$1975\pm27$	&	 \cite{Alametal2017} \\
	&	0.61	&	$(2283\pm32)r_d/r_{d, {\rm fid}}$	&	$2280\pm32$	&	\\ \hline
$D_V$	&	1.52	&	$(3843\pm147)r_d/r_{d, {\rm fid}}$	&	$3838\pm147$	 &	 \cite{Ataetal2018} \\ \hline
$D_M$	&	2.33	&	$(37.77\pm2.13)r_d$	&	$5575\pm314$	&	 \cite{Bautistaetal2017} \\ \hline
\end{tabular}
\end{center}
\end{table}

From Table \ref{tab:BAO}, the vector of observed data is $V_{\rm obs}\equiv[D_V(0.106), D_V(0.15), D_M(0.38), D_M(0.51), D_M(0.61), D_V(1.52), \\ D_M(2.33)]$. In addition, with eqn.\ (\ref{eq:dpdm}) and the smoothed $H(z)$ function, we
can construct the corresponding ``model prediction'' vector $V_{\rm th}(H_0,\Omega_{K0})$ which depends on the values of $H_0$ and $\Omega_{K0}$. Here, we use the smoothed $H(z)$ function derived from Sample 1{\_}0 to be able to make use of the
BAO distance measurement at $z=2.33$. (Sample 1{\_}0 is comprised of the 31
cosmic chronometric $H(z)$ measurements as well as the two highest $z$ BAO
Ly$\alpha$ ones.) The two-dimensional log likelihood function is then
given by
\begin{equation}\label{eq:VC}
    \ln[{\mathcal{L}}(\Omega_{K0}, H_0)]=-0.5(V_{\rm obs}-V_{\rm th})C^{-1}(V_{\rm obs}-V_{\rm th})^T,
\end{equation}
where the covariance matrix of the observed BAO distances is
\begin{equation}\label{eq:matrix}
C =
 \left(
   \begin{array}{ccccccc}
729	&	0	&	0	&	0	&	0	&	0	&	0	\\
0	&	616	&	0	&	0	&	0	&	0	&	0	\\
0	&	0	&	483	&	294	&	140	&	0	&	0	\\
0	&	0	&	294	&	727	&	441	&	0	&	0	\\
0	&	0	&	140	&	440	&	1022	&	0	&	0	\\
0	&	0	&	0	&	0	&	0	&	21556	&	0	\\
0	&	0	&	0	&	0	&	0	&	0	&	98840	\\
   \end{array}
 \right).
\end{equation}

\begin{figure}
  \includegraphics[width=0.5\textwidth]{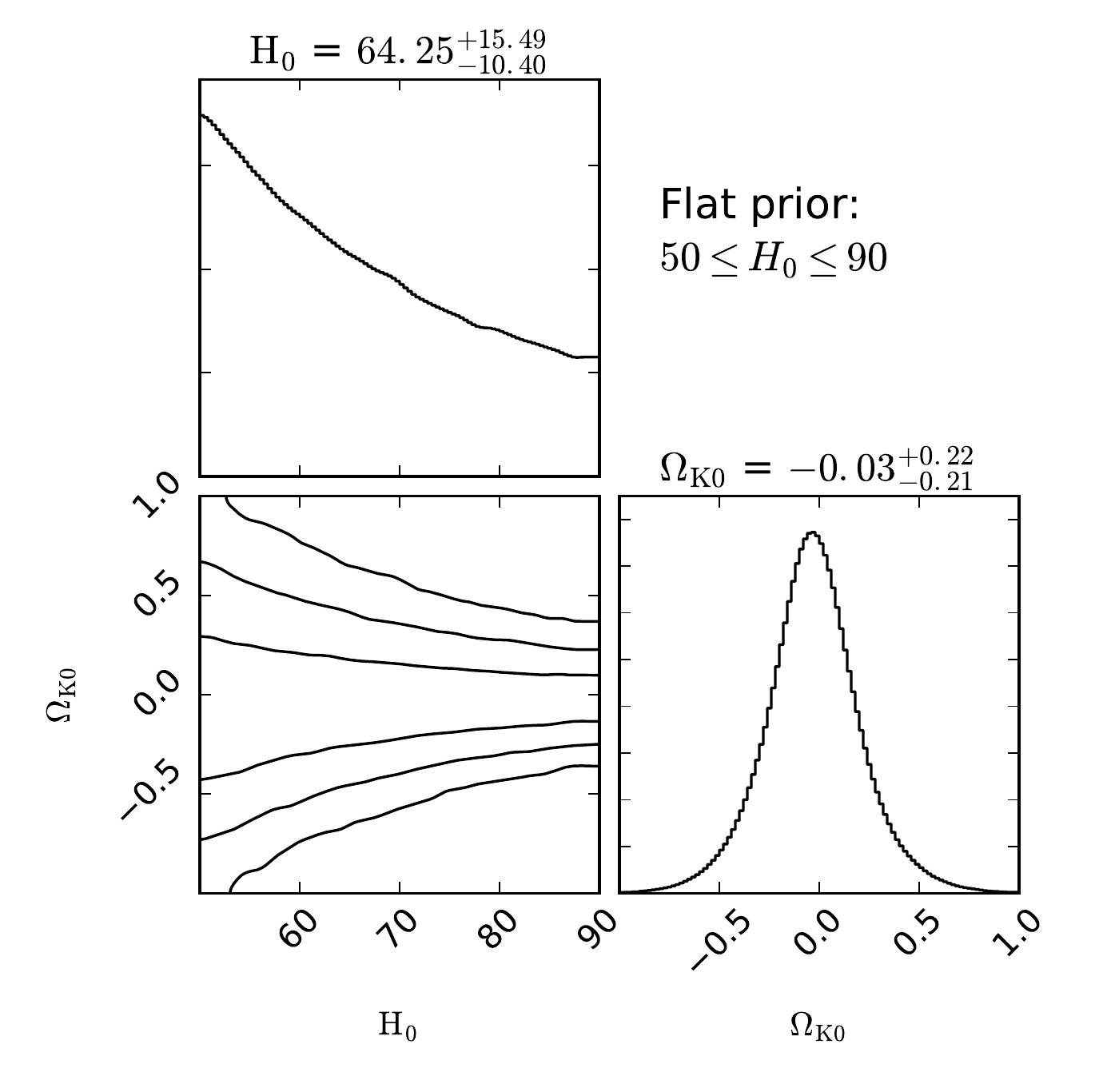}
  \includegraphics[width=0.5\textwidth]{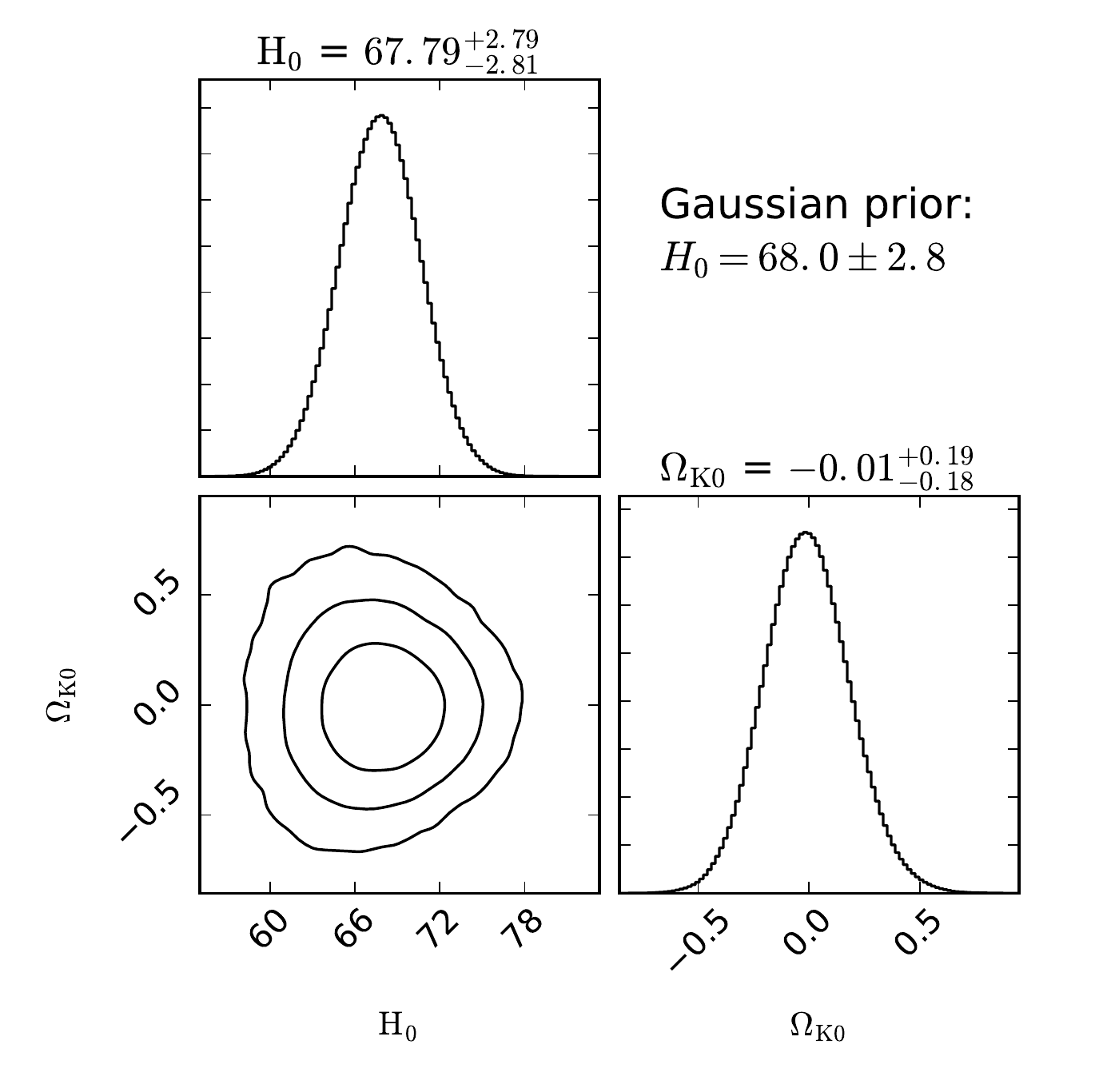}\\
  \includegraphics[width=0.5\textwidth]{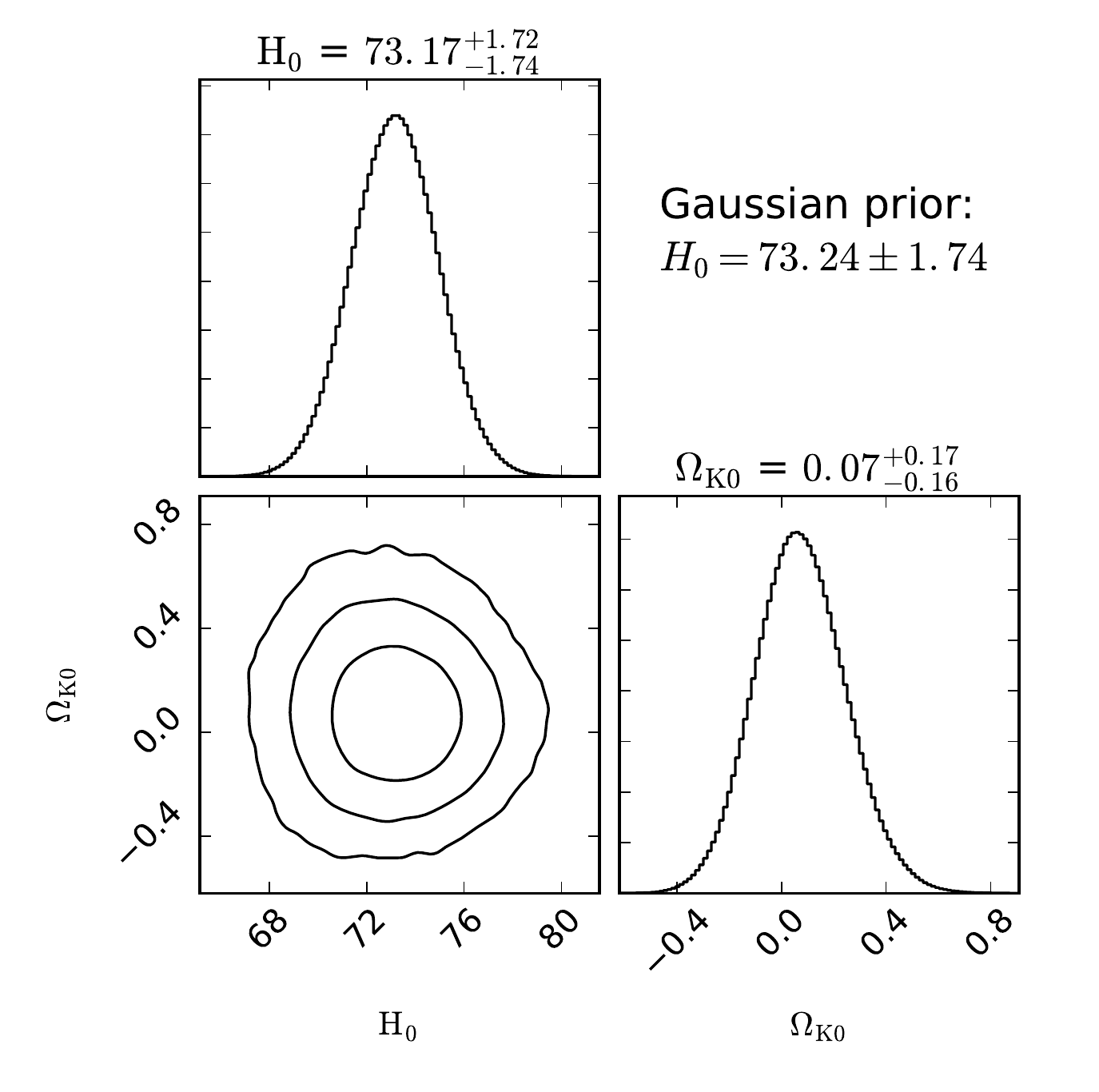}
  \caption{Constraints on $\Omega_{K0}$ for the three different $H_0$ priors. The left top panels present the flat prior non-zero over only $50\,\rm km/s/Mpc\leq H_0\leq90\,\rm km/s/Mpc$, the right top panels present results for the gaussian prior corresponding to $H_0=68\pm2.8\,\rm km/s/Mpc$, while the last one is for the gaussian prior with $H_0=73.24\pm1.74\,\rm km/s/Mpc$.}\label{fig:Omega_K}
\end{figure}

We use the open-source Markov chain Monte Carlo Python package $emcee$
\citep{Foreman-Mackeyetal2013} for our analyses. The two-dimensional
likelihood function depends on $\Omega_{K0}$ and $H_0$. To derive
one-dimensional
$H_0$ and $\Omega_{K0}$ likelihoods and limits we need to assume priors. We
use three priors for $H_0$, a flat prior, non-zero over $50-90\,\rm km/s/Mpc$,
and two gaussian priors with $H_0 = 68.0\pm2.8\rm\,km/s/Mpc$
\citep{ChenRatra2011a} and $H_0 = 73.24\pm1.74\rm\,km/s/Mpc$
\citep{Riessetal2016}. The $\Omega_{K0}$ prior is chosen to be flat and non-zero
over $-1\leq\Omega_{K0}\leq1$. Results are shown in Figure\ \ref{fig:Omega_K} for
the gaussian covariance function case; the other two covariance functions lead
to very similar results.

From Figure\ \ref{fig:Omega_K}, we see that $\Omega_{K0}$ is constrained to
$-0.03\pm0.21$, $-0.01\pm0.19$, and $0.07\pm0.17$ when using the flat prior
non-zero over the $H_0$ range $50-90\,\rm km/s/Mpc$, and the two gaussian
priors with $H_0 = 68\pm2.8\,\rm km/s/Mpc$ and
$H_0 = 73.24\pm1.74\,\rm km/s/Mpc$ respectively.
These results are consistent with the results of earlier work that use $H(z)$,
BAO, and other non CMB anisotropy data to measure $\Omega_{K0}$ \citep{Farooqetal2015, YuWang2016, Farooqetal2017, Lietal2016, WeiWu2017, Xiaetal2017, Ranaetal2017, Wangetal2017a, Mitraetal2017}.
Although these results are mostly consistent with a flat universe, the
uncertainties are large and more precise measurements of $H(z)$ and the angular
diameter distance at higher redshift are needed for tighter constraints on
$\Omega_{K0}$ and to check the findings of \citet{Oobaetal2017a, Oobaetal2017b, Oobaetal2017c}, and \citet{ParkRatra2018}.

\section{Conclusion}\label{sec:summary}

We compile a complete collection of currently available, reliable
$H(z)$ data. Of these 36 $H(z)$ data points, 31 are measured by the cosmic
chronometric technique, while 5 come from BAO observations. We use this
compilation with the GP method to determine a continuous $H(z)$ function,
from which we measure $H_0$, $z_{\rm da}$, and $\Omega_{K0}$ (in combination
with BAO distance-redshift data) and use to test the
flat-$\Lambda$CDM model. While there has been a lot of earlier work on these
issues, we have for the first time accounted for all known correlations
between data points that have previously been ignored. We have also
organized our $H(z)$ data into 12 different samples to check for potential
effects caused by the measurement or data reduction technique used or by
the value assumed for $H_0$.

Averaging across the samples we use, we find
$H_0\sim 67 \pm 4\,\rm km/s/Mpc$, more consistent with the recent lower
values determined using a variety of techniques.

In most samples we consider, we see a cosmological deceleration-acceleration
transition at 2$\sigma$ significance, with the data sample transition
redshifts varying over $0.33<z_{\rm da}<1.0$ at 1$\sigma$ significance.
This is significantly broader than the $z_{\rm da}$ range determined using
cosmological model templates. The reason for this disagreement might be
that the GP method allows a less steep expansion history at $z>1$, than do
the cosmological model templates, which make the $\ddot{a}$ function
flatter. This needs to be studied more carefully when more precise
$H(z)$ measurements become available.

 We find that the flat-$\Lambda$CDM model is consistent with the $H(z)$ data to
a $z$ of 1.5 to 2.0 depending on the sample considered, with 2$\sigma$
deviations from flat-$\Lambda$CDM above this redshift range. This also needs
to be more carefully examined with future higher-quality $H(z)$ data.

Using the continuous $H(z)$ with BAO distance-redshift observations, a
representative constraint on the current spatial curvature density parameter is
$\Omega_{K0}=-0.03\pm0.21$. This is consistent with a flat universe, but the
large error bar does not rule out small values of spatial curvature
that might be of interest \citep{Oobaetal2017a, Oobaetal2017b}. Higher
quality measurements of $H(z)$ and $D_M(z)$ at higher redshift are
necessary to tighten this constraint, and these are likely to become available
soon.

\section{Acknowledgments}\label{sec:ack}
This work is supported by  the National Basic Research Program of
China (973 Program, grant No. 2014CB845800), the National Natural
Science Foundation of China (grants 11422325 and 11373022), the
Excellent Youth Foundation of Jiangsu Province (BK20140016), and
by DOE grant DE-SC0011840. Yu, H. also thanks the support of China Scholarship Council for studying abroad.

\end{document}